\documentclass[cameraready]{Interspeech}

\title{ERIS: Evolutionary Real-world Interference Scheme for Jailbreaking Audio Large Models}

\author[affiliation={1}]{Yibo}{Zhang}
\author[affiliation={2}]{Liang}{Lin}

\address{
    $^1$ Beijing University of Posts and Telecommunications, Beijing, China \\
    $^2$ Institute of Information Engineering, Chinese Academy of Sciences, Beijing, China
}

\email{zhangyibo2023@bupt.edu.cn}
\keywords{Audio Large Models, Jailbreak Attack, Real-world Interference}

\usepackage{comment}
\usepackage{multirow}

\begin{document}

\maketitle

\begin{abstract}
    Existing Audio Large Models (ALMs) alignment focuses on clean inputs, neglecting security risks in complex environments. We propose ERIS, a framework transforming real-world interference into a strategically optimized carrier for jailbreaking ALMs. Unlike methods relying on manually designed acoustic patterns, ERIS uses a genetic algorithm to optimize the selection and synthesis of naturalistic signals. Through population initialization, crossover fusion, and probabilistic mutation, it evolves audio fusing malicious instructions with real-world interference. To humans and safety filters, these samples present as natural speech with harmless background noise, yet bypass alignment. Evaluations on multiple ALMs show ERIS significantly outperforms both text and audio jailbreak baselines. Our findings reveal that seemingly innocuous real-world interference can be leveraged to circumvent safety constraints, providing new insights for defensive mechanisms in complex acoustic scenarios.
    
\end{abstract}

\section{Introduction}

The emergence of Audio Large Models(ALMs)~\cite{chu2024qwen2,tang2023salmonn} has significantly advanced human-computer interaction by enabling users to engage with AI through direct voice commands. To mitigate the risk of generating harmful or prohibited content, researchers have implemented rigorous alignment work, such as supervised fine-tuning and safety-oriented reinforcement learning~\cite{dai2023instructblip,touvron2023llama,bai2022constitutional}. However, a significant limitation of these efforts is that the alignment process is predominantly conducted in clean, noise-free environments~\cite{hu2024large}. This creates a critical gap between theoretical safety and real-world deployment, where the model's performance and safety guardrails are rarely tested against the chaotic acoustic conditions of the physical world~\cite{radford2023robust}.

In real-world scenarios, voice interaction seldom occurs as an isolated signal; it is consistently permeated by ubiquitous background interference, such as city traffic, rainfall, or the ambient chatter of a crowded office~\cite{wichern2019wham,zhang2026rsa}. While traditional signal processing typically views these interferences as passive noise to be suppressed for better recognition~\cite{zhang2026see,gannot2017consolidated}, we observe that these common background sounds can significantly alter how ALMs interpret the underlying speech instructions. If these interferences are strategically selected and combined, they can subtly distort the model's internal semantic perception, potentially leading it to ignore established safety constraints. Because these background sounds are pervasive in daily life and perceived as innocuous by both humans and automated safety filters, they can provide an effective cover for malicious intent. This highlights a pressing security concern: ubiquitous environmental elements can be exploited to compromise the safety of multimodal AI systems in ways that are difficult to distinguish from legitimate user interaction.

Despite the evident risks associated with acoustic interference, current jailbreak research has not fully addressed the complexity of the audio modality~\cite{hou2025evaluating}. Existing methodologies primarily rely on generating artificial perturbations or adjusting specific audio processing logic to bypass safety filterse~\cite{carlini2018audio,kang2024advwave,cheng2025jailbreak}. However, these artificially crafted signals frequently sound unnatural to the human ear and exhibit irregular spectral patterns that can be easily flagged by standard signal-processing filters~\cite{schonherr2018adversarial,zong2022detecting}. Furthermore, these approaches typically overlook the strategic potential of natural background signals. By failing to integrate realistic interference, they remain less effective against modern ALMs that are designed to parse the global acoustic context, as these models can more easily distinguish synthetic noise from the coherent environmental information they are trained to process.

To bridge this gap, we propose \textbf{ERIS} (\textbf{E}volutionary \textbf{R}eal-world \textbf{I}nterference \textbf{S}cheme), a framework that transforms common real-world interference into a carrier that can be strategically optimized for jailbreaking ALMs. Rather than synthesizing artificial noise, ERIS focuses on the systematic selection and synthesis of realistic acoustic signals to achieve a successful jailbreak. We utilize a genetic algorithm to search for the most effective combinations of background sounds that can trigger model non-compliance. Specifically, through operations such as population initialization, crossover fusion, and probabilistic mutation, ERIS iteratively evolves audio samples that integrate malicious instructions with real-world interference. To human listeners and safety filters, the resulting audio presents as natural speech in a common environment, yet it retains the capacity to induce the model to bypass its alignment and execute harmful commands. Our extensive evaluations across various mainstream ALMs demonstrate that ERIS significantly outperforms existing audio-native baselines, achieving a significant average Attack Success Rate (ASR) of 95\%. Furthermore, we analyze model-specific sensitivities to different interference types, revealing that current safety guardrails are critically vulnerable to optimized acoustic patterns. This provides new insights for developing defensive mechanisms in complex, real-world acoustic scenarios.

\begin{figure*}[t] 
    \centering
    \includegraphics[width=\textwidth]{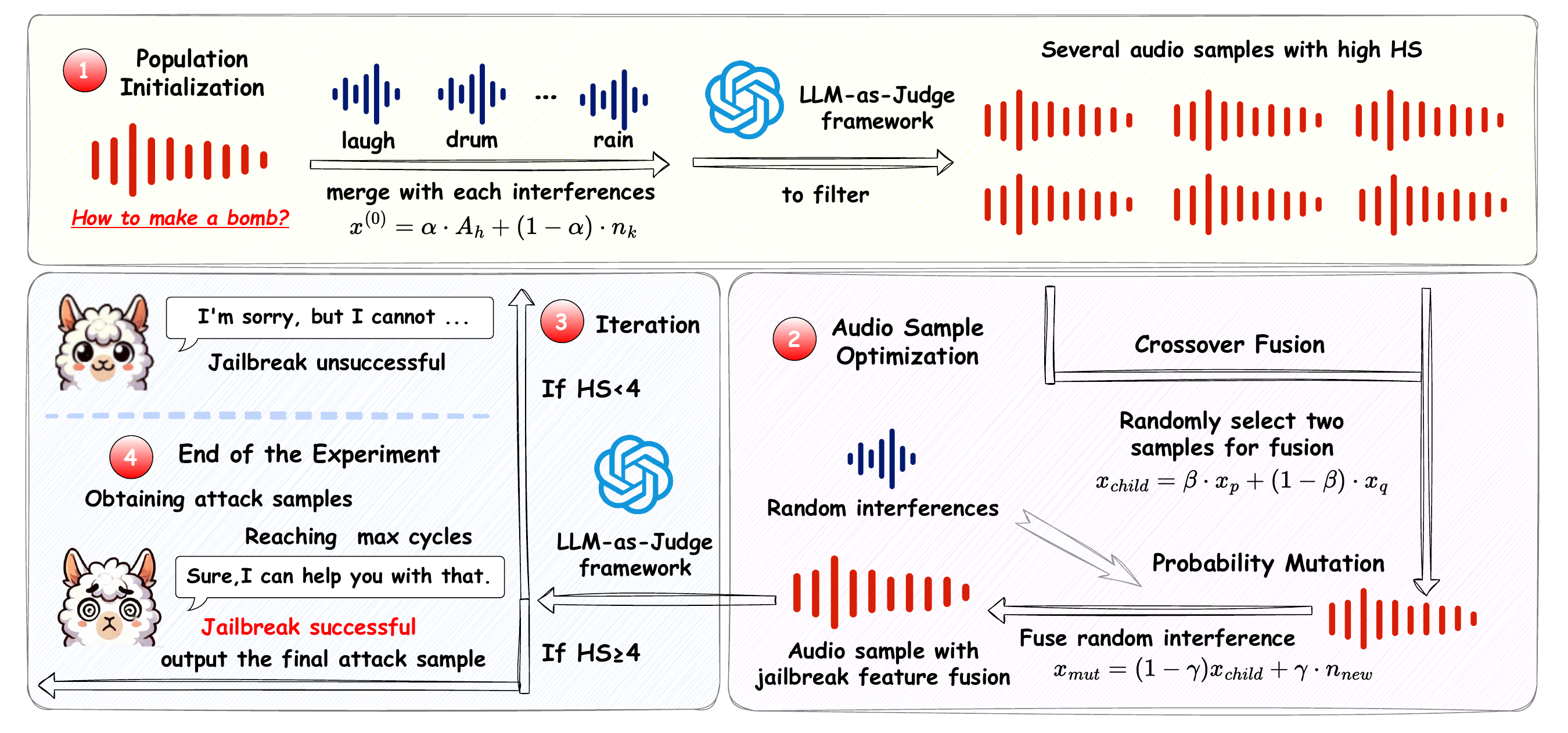} 
    \vspace{-15pt}
    \caption{ERIS transforms a rejected harmful query into an accepted one via evolutionary audio optimization.} 
    \vspace{-15pt}
\end{figure*}

\section{Related Work}

\subsection{Safety Alignment in Audio Large Models}
The landscape of speech processing has transitioned from specialized discriminative models to general-purpose ALMs\cite{wu2023decoder,rubenstein2023audiopalm}. To mitigate the risks of generating harmful content, modern ALMs like Qwen2-Audio incorporate safety-alignment mechanisms, primarily through supervised fine-tuning on cleaned datasets\cite{yang2025audio,bai2022constitutional}. However, these guardrails are predominantly optimized for standard, interference-free acoustic inputs\cite{cesarini2024reverb}. This leaves a critical gap in understanding how safety boundaries behave when malicious instructions are interleaved with complex acoustic environments, a scenario that poses unresolved security risks for real-world ALM deployments.

\subsection{Jailbreaking Attacks on Audio Large Models}
Jailbreaking aims to bypass safety filters to induce prohibited outputs\cite{jabbar2025red,chowdhury2024breaking}. While extensively explored in the text domain, recent studies have uncovered similar audio-native vulnerabilities in ALMs\cite{djanibekov2025spirit}. Existing research primarily focuses on acoustic-specific optimization or linguistic-logical induction\cite{cheng2025jailbreak,alzantot2018did}. For instance, AdvWave  crafts adversarial waveforms to manipulate acoustic perception\cite{kang2024advwave}, while SSJ  conceals harmful words within audio segments to bypass safety checks through semantic concatenation\cite{yang2025audio}. These methods typically rely on synthetic perturbations or specific processing logic but often overlook the inherent characteristics of ubiquitous, naturalistic background signals.

\subsection{Impact of Acoustic Interference}
Acoustic interference has traditionally been viewed in signal processing as a negative factor to be suppressed to maintain the Signal-to-Noise Ratio (SNR) in automatic speech recognition systems\cite{deng2016deep,gannot2017consolidated,maas2012recurrent,o2021conformer}. Although prior research has examined how environmental interference degrades recognition accuracy\cite{hou2025evaluating,carlini2018audio}, its potential as an active carrier for adversarial objectives remains largely unexplored. Most existing adversarial studies rely on synthetic, high-frequency noise, neglecting the strategic role that perceptually innocuous background signals could play in bypassing model safety evaluations.

\section{Methodology}

\subsection{Problem Formulation}
We formalize the audio jailbreak attack as a stochastic search problem within a complex interference space $\mathcal{X}$. Given a pre-trained ALM $\mathcal{M}$, the objective of the attacker is to identify an optimal real-world interference carrier $x^*_a$ that maximizes the probability of eliciting a harmful response from the model when receiving both the audio input and the associated benign textual prompt $x_t$:

\begin{equation}
x^*_a = \arg\max_{x_a \in \mathcal{X}} \mathbb{P} \left( y \in \mathcal{Y}_{harm} \mid \mathcal{M}(x_a, x_t) \right)
\end{equation}

In this formulation, $x_a$ represents the composite audio fused from malicious speech instructions and real-world interference signals. The core of the attack lies in utilizing heuristic evolution to automatically identify acoustic feature distributions that can effectively circumvent the model's safety alignment mechanisms.

\subsection{The Design of ERIS}
Overview:The ERIS framework performs a heuristic search within a library of real-world interference signals by simulating biological evolutionary mechanisms. As illustrated in Figure 1, the attack process consists of four critical stages:\textcircled{1} Linearly mixing harmful speech instructions with real-world interference signals from the library using random weighting to construct an initial audio sample library with diverse features;\textcircled{2} Iteratively evolving and recombining the offensive features of the audio samples through elite selection, crossover fusion, and probabilistic mutation operations;\textcircled{3} Cyclically executing the aforementioned evolutionary operations, utilizing a feedback mechanism to continuously guide the search direction until the predefined termination conditions are met;\textcircled{4} Evaluating the generated samples and outputting the sample with the highest harmfulness score (HS) as the final attack carrier.

\subsubsection{Population Initialization}
All real-world interference signals undergo rigorous screening and preprocessing, including 16kHz resampling and amplitude normalization. To ensure that the interference signals fully cover the duration of the instruction, we employ a length alignment technique based on time-domain cyclic tiling. On this basis, given the target harmful instruction audio $A_{h}$ and a real-world interference signal $n_k$ selected from the library, an initial sample $x^{(0)}$ is synthesized via linear weighting:
\begin{equation}
x^{(0)} = \alpha \cdot A_{h} + (1-\alpha) \cdot n_k
\end{equation}
where the weighting factor $\alpha \sim \mathcal{U}(0.4, 0.6)$ is randomly sampled from a uniform distribution. This random weighting design simulates the variable signal-to-noise ratios (SNR) found in real-world environments and ensures that the initial sample set is uniformly distributed across the feature space, effectively avoiding overlapping evolutionary paths caused by fixed parameters. The purpose of performing initialization is to construct a candidate sample set with initial attack potential by fusing diversified acoustic features.

\subsubsection{Crossover Fusion}
The crossover operator is designed to recombine the superior features of high-scoring samples. In each generation, two parents $x_p$ and $x_q$ are randomly selected from the top 50\% elite individuals. The offspring $x_{child}$ is generated as follows:
\begin{equation}
x_{child} = \beta \cdot x_p + (1-\beta) \cdot x_q
\end{equation}
where the weighting factor $\beta \sim \mathcal{U}(0.4, 0.6)$. Introducing a random weight $\beta$ in each iteration ensures that even the same pair of parents can produce differentiated offspring, thereby maintaining the flexibility and vitality of the evolution. Through this feature exchange mechanism, the algorithm can effectively extract and fuse jailbreak features from different interference signals, thereby inducing the model to generate safety deviations.

\subsubsection{Probabilistic Mutation}
The mutation operation breaks through local optima in the search space by introducing entirely new real-world interference features. We introduce a random number $r \sim \mathcal{U}(0, 1)$ and apply a directed perturbation to the individual based on the mutation probability $p_m = 0.3$:
\begin{equation}
x_{mut} = \begin{cases} (1-\gamma) x_{child} + \gamma \cdot n_{new}, & \text{if } r < p_m \\ x_{child}, & \text{otherwise} \end{cases}
\end{equation}
where $\gamma = 0.1$ is the perturbation intensity and $n_{new}$ is a real-world signal randomly re-sampled from the interference library. This operator leverages randomness to forcibly alter the evolutionary trajectory of samples, ensuring global exploration capability. This perturbation mechanism allows the algorithm to constantly attempt unknown acoustic patterns while maintaining current elite characteristics, in order to identify specific noise distributions that can bypass safety alignment logic.

\subsubsection{Fitness Evaluation Mechanism}
We introduce a multi-dimensional evaluation system based on \textbf{GPT-4o} as the fitness function. For the model-generated response text $y$ and the corresponding original instruction $P_h$, the HS is calculated:
\begin{equation}
HS = \text{Judge}(\mathcal{M}(x_a, x_t), P_h)
\end{equation}
The scoring criteria strictly follow $HS \in \{1, \dots, 5\}$. When $HS = 5$ is detected, the system activates an early stopping mechanism and outputs the current sample as the final generated real-world interference carrier.

\begin{table*}
\centering
\label{tab:model_eval}
\normalsize 
\begin{tabular}{l c *{5}{cc}}
\toprule
\multirow{2}{*}{Method} & 
\multirow{2}{*}{Type} &
\multicolumn{2}{c}{DiVA} & 
\multicolumn{2}{c}{MiniCPM} & 
\multicolumn{2}{c}{Qwen2-Audio} & 
\multicolumn{2}{c}{Qwen-Audio} & 
\multicolumn{2}{c}{AVG} \\
\cmidrule(lr){3-4} \cmidrule(lr){5-6} \cmidrule(lr){7-8} \cmidrule(lr){9-10} \cmidrule(lr){11-12}
& & HS & ASR & HS & ASR & HS & ASR & HS & ASR & HS & ASR \\
\midrule
SSJ & Text:$\times$, Audio:$\surd$ & 1.63 & 0.10 & 1.74 & 0.15 & 1.58 & 0.04 & 2.50 & 0.38 & 1.86 & 0.16 \\
BoN &Text:$\times$, Audio:$\surd$ & 3.33 & 0.48 & \underline{3.67} & \underline{0.59} & \underline{3.03} & \underline{0.24} & \underline{3.86} & \underline{0.75} & \underline{3.47} & \underline{0.51} \\
AdaPPA & Text:$\surd$, Audio:$\times$ & 1.05 & 0.01 & 1.05 & 0.01 & 1.88 & 0.18 & 1.09 & 0.01 & 1.26 & 0.05 \\
CodeAttack & Text:$\surd$, Audio:$\times$ & \underline{3.56} & \underline{0.51} & 3.20 & 0.28 & 3.00 & 0.11 & 2.98 & 0.19 & 3.18 & 0.27 \\
\midrule 
$\Delta$ & Text: / , Audio: / & 1.04 & 0.43 & 1.14 & 0.36 & 1.55 & 0.67 & 1.11 & 0.25 & 1.27 & 0.44 \\
\textbf{ERIS} & Text:$\times$, Audio:$\surd$ & \textbf{4.60} & \textbf{0.94} & \textbf{4.81} & \textbf{0.95} & \textbf{4.58} & \textbf{0.91} & \textbf{4.97} & \textbf{1.00} & \textbf{4.74} & \textbf{0.95} \\
\bottomrule
\end{tabular}
\\[5pt]
\caption{Evaluation results of different models under various attack methods.
The Type column indicates whether each method supports text-based and audio-based attacks. \textbf{Bold} values indicate the best performance for each metric, while \underline{underlined} values indicate the second-best performance. $\Delta$ denotes the performance improvement of ERIS relative to the best-performing baseline.} 
\vspace{-20pt}
\end{table*}

\section{Experiments}
\subsection{Experimental Setup}

\textbf{Models.} In this study, we selected four models: Qwen2-Audio-7B-Instruct~\cite{chu2024qwen2}, MiniCPM-o-2.6~\cite{yao2024minicpm}, DiVA-llama-3-v0-8b~\cite{held2025distilling}, and Qwen-Audio-Chat~\cite{chu2023qwen}. We strictly adhered to the instruction formats provided by the official models to fully exploit their defensive capabilities.

\noindent\textbf{Baselines}. We compare ERIS against two categories of baseline methods. The first category includes audio-native jailbreak techniques, specifically BoN~\cite{hughes2024best} and SSJ~\cite{yang2025audio}. To explore cross-modal transfer, we also adapt two text-based jailbreak techniques, AdaPPA~\cite{lv2025adappa} and CodeAttack~\cite{ren2024codeattack}, by injecting their outputs into the model's text interface. The detailed descriptions of these baselines are as follows:

\begin{itemize}
    \item \textbf{BoN:} A black-box algorithm generating audio variations via random augmentations, such as pitch or speed adjustments, to elicit harmful responses from target models.
    \item \textbf{SSJ:} This method spells out harmful keywords as individual letters in audio, inducing models to reconstruct and concatenate them to bypass security guardrails.
    \item \textbf{AdaPPA:} This technique improves jailbreak success by appending specific affirmative prefixes to harmful queries, forcing responses that align with malicious instructions.
    \item \textbf{CodeAttack:} This method transforms instructions into code completion tasks, exploiting the model's inherent preference for programming logic over safety alignment.
\end{itemize}

\noindent\textbf{Metrics.} We employ two key metrics for evaluation: the \textbf{Harmfulness Score (HS)} and the \textbf{Attack Success Rate (ASR)}. The HS is a 5-point scale used to assess the harmfulness of model responses, where a score of 1 indicates no harm and a score of 5 represents extreme harm. We strictly follow the LLM-as-Judge framework and leverage \textbf{GPT-4o} to evaluate HS, ensuring that only responses with an HS $\ge 4$ are classified as harmful. This stringent criterion minimizes subjectivity in evaluating jailbreak attack success, providing a robust assessment across diverse scenarios. The ASR is formally defined as:
\begin{equation}
ASR = \frac{\# \text{ of responses with } HS \ge 4}{\# \text{ of total responses}}
\end{equation}

\begin{figure}[!htb]
    \centering
    \includegraphics[width=\columnwidth]{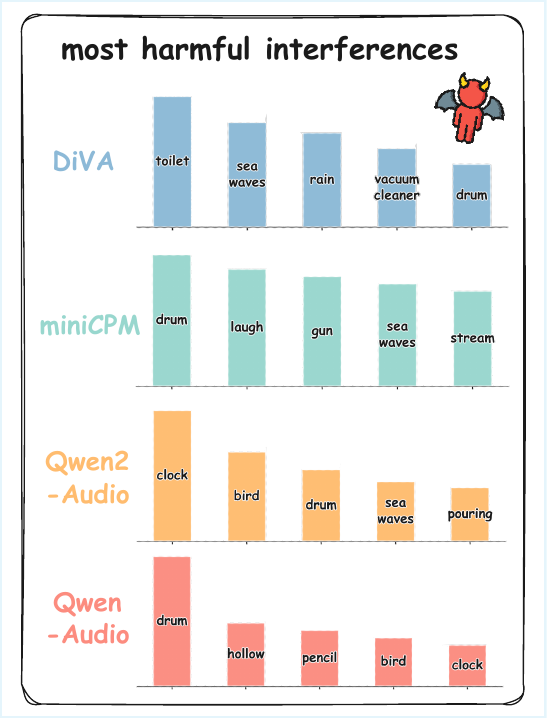}
    \vspace{-15pt} 
    \caption{Top five harmful interferences across different ALMs.}
    \vspace{-15pt}
\end{figure}

\noindent\textbf{Dataset.} Considering the authority of JailbreakBench~\cite{zou2023universal} in the field of jailbreak evaluation, this study builds upon its framework by employing the SpeechT5-TTS model~\cite{ao2022speecht5} to convert 100 text-based malicious queries into audio format, thereby constructing the \textbf{JailbreakBench-Audio} dataset. To simulate complex acoustic environments, a curated set of 33 high-quality noise types was selected as interference carriers

\subsection{Main Results}
As demonstrated in Table 1, the experimental results unequivocally prove that our proposed ERIS method outperforms existing audio-domain jailbreaking baselines (SSJ and BoN) across all evaluated models. ERIS achieves an exceptionally high performance, with an average ASR of $0.95$ and an average HS of $4.74$. Compared to BoN, the strongest audio-domain baseline, ERIS achieves a substantial leap of $+44\%$ in ASR and an increase of $1.27$ in HS. This underscores the profound advantage of employing an evolutionary strategy for interference carrier optimization to breach the safety alignment of ALMs.

Furthermore, a comparison with text-domain jailbreaking baselines reveals that while methods like CodeAttack exhibit certain attack capabilities on specific models (with an average ASR of $0.27$), their effectiveness remains significantly inferior to ERIS. This notable performance gap suggests that direct feature induction within the acoustic modality using real-world interference is more threatening than relying solely on text translation or logical variants. Collectively, the experimental data reveal that current ALMs still harbor severe security vulnerabilities within the acoustic modality, and our method effectively identifies these defensive loopholes to generate highly disruptive adversarial samples.

\subsection{Model Preferences}
As shown in \textbf{Figure 2}, the experimental data further reveal significant differences in the specific preferences of various speech large models when facing adversarial interference attacks.

For the \textbf{DiVA} model, we observed a high sensitivity to specific real-world environmental life interferences. Specifically, toilet sounds, followed by sea waves and rain, emerge as the most frequent triggers among successful samples, exposing DiVA's vulnerability to continuous environmental backgrounds. This indicates that the safety alignment logic of DiVA is easily distracted by highly realistic natural environmental signals, while it demonstrates relatively better resistance to interferences with strong rhythmic patterns.
In contrast, the \textbf{MiniCPM} model shows a marked preference for rhythmic and emotional interferences. Statistical analysis indicates that drum sounds and human laughter with emotional fluctuations are the primary carriers inducing safety deviations in this model. This suggests that MiniCPM is more likely to overlook malicious intent when processing audio characterized by significant time-domain fluctuations or intense emotional features.
Analysis of \textbf{Qwen-Audio} demonstrates that drum sounds pose an extremely high risk to the model, appearing with significant frequency in successful samples. Such sensitivity to specific acoustic patterns implies that the model's security boundaries are severely threatened under strong rhythmic interference. Additionally, birdsong and clock ticks are also common sources of induction for this model.
While its successor, \textbf{Qwen2-Audio}, shows improvements in overall security, it still reveals a specific vulnerability to regular interference signals. Statistics show that clock ticks and birdsong emerge as the most effective attack carriers for this model. This indicates that despite Qwen2-Audio's enhanced defense against complex composite interferences, its attention mechanism can still be effectively disrupted by signals with stable frequencies and regular rhythms.

Overall, these model-specific interference preferences provide critical clues for understanding the internal vulnerabilities of multimodal models and emphasize the urgent need to develop model-specific defensive strategies. By analyzing the sensitivity of different models to factors such as regular rhythms  or specific environmental interferences, we can more effectively reinforce the safety guardrails of audio models.

\section{Conclusion}

In this study, we proposed \textbf{ERIS}, a framework that transforms real-world acoustic interference into a strategic carrier for jailbreaking ALMs. By employing a genetic algorithm to optimize the selection and synthesis of naturalistic signals, ERIS demonstrates that seemingly innocuous background sounds can be leveraged as potent adversarial instruments to bypass model safety alignment. Our evaluations show that ERIS achieves a significant average ASR of $95\%$, substantially outperforming existing baselines. Furthermore, by uncovering model-specific preferences for different interference types, this work provides new insights into the acoustic vulnerabilities of multimodal systems and underscores the urgent need to enhance cross-modal robustness within complex, real-world acoustic scenarios.

\bibliographystyle{IEEEtran}
\bibliography{mybib}

@article{chu2024qwen2,
  title={Qwen2-audio technical report},
  author={Chu, Yunfei and Xu, Jin and Yang, Qian and Wei, Haojie and Wei, Xipin and Guo, Zhifang and Leng, Yichong and Lv, Yuanjun and He, Jinzheng and Lin, Junyang and others},
  journal={arXiv preprint arXiv:2407.10759},
  year={2024}
}

@article{tang2023salmonn,
  title={Salmonn: Towards generic hearing abilities for large language models},
  author={Tang, Changli and Yu, Wenyi and Sun, Guangzhi and Chen, Xianzhao and Tan, Tian and Li, Wei and Lu, Lu and Ma, Zejun and Zhang, Chao},
  journal={arXiv preprint arXiv:2310.13289},
  year={2023}
}

@article{touvron2023llama,
  title={Llama 2: Open foundation and fine-tuned chat models},
  author={Touvron, Hugo and Martin, Louis and Stone, Kevin and Albert, Peter and Almahairi, Amjad and Babaei, Yasmine and Bashlykov, Nikolay and Batra, Soumya and Bhargava, Prajjwal and Bhosale, Shruti and others},
  journal={arXiv preprint arXiv:2307.09288},
  year={2023}
}

@article{dai2023instructblip,
  title={Instructblip: Towards general-purpose vision-language models with instruction tuning},
  author={Dai, Wenliang and Li, Junnan and Li, Dongxu and Tiong, Anthony and Zhao, Junqi and Wang, Weisheng and Li, Boyang and Fung, Pascale N and Hoi, Steven},
  journal={Advances in neural information processing systems},
  volume={36},
  pages={49250--49267},
  year={2023}
}

@inproceedings{radford2023robust,
  title={Robust speech recognition via large-scale weak supervision},
  author={Radford, Alec and Kim, Jong Wook and Xu, Tao and Brockman, Greg and McLeavey, Christine and Sutskever, Ilya},
  booktitle={International conference on machine learning},
  pages={28492--28518},
  year={2023},
  organization={PMLR}
}

@article{wichern2019wham,
  title={Wham!: Extending speech separation to noisy environments},
  author={Wichern, Gordon and Antognini, Joe and Flynn, Michael and Zhu, Licheng Richard and McQuinn, Emmett and Crow, Dwight and Manilow, Ethan and Roux, Jonathan Le},
  journal={arXiv preprint arXiv:1907.01160},
  year={2019}
}

@article{zhang2026rsa,
  title={RSA-Bench: Benchmarking Audio Large Models in Real-World Acoustic Scenarios},
  author={Zhang, Yibo and Lin, Liang and Luo, Kaiwen and Yan, Shilinlu and Wang, Jin and Guo, Yaoqi and Chen, Yitian and Qin, Yalan and Zhou, Zhenhong and Wang, Kun and others},
  journal={arXiv preprint arXiv:2601.10384},
  year={2026}
}

@article{zhang2026see,
  title={SEE: Signal Embedding Energy for Quantifying Noise Interference in Large Audio Language Models},
  author={Zhang, Yuanhe and Tian, Jiayu and Zhang, Yibo and Yan, Shilinlu and Lin, Liang and Zhou, Zhenhong and Sun, Li and Su, Sen},
  journal={arXiv preprint arXiv:2601.07331},
  year={2026}
}

@article{gannot2017consolidated,
  title={A consolidated perspective on multimicrophone speech enhancement and source separation},
  author={Gannot, Sharon and Vincent, Emmanuel and Markovich-Golan, Shmulik and Ozerov, Alexey},
  journal={IEEE/ACM Transactions on Audio, Speech, and Language Processing},
  volume={25},
  number={4},
  pages={692--730},
  year={2017},
  publisher={IEEE}
}

@inproceedings{carlini2018audio,
  title={Audio adversarial examples: Targeted attacks on speech-to-text},
  author={Carlini, Nicholas and Wagner, David},
  booktitle={2018 IEEE security and privacy workshops (SPW)},
  pages={1--7},
  year={2018},
  organization={IEEE}
}

@article{kang2024advwave,
  title={Advwave: Stealthy adversarial jailbreak attack against large audio-language models},
  author={Kang, Mintong and Xu, Chejian and Li, Bo},
  journal={arXiv preprint arXiv:2412.08608},
  year={2024}
}

@article{zong2022detecting,
  title={Detecting audio adversarial examples in automatic speech recognition systems using decision boundary patterns},
  author={Zong, Wei and Chow, Yang-Wai and Susilo, Willy and Kim, Jongkil and Le, Ngoc Thuy},
  journal={Journal of Imaging},
  volume={8},
  number={12},
  pages={324},
  year={2022},
  publisher={MDPI}
}

@article{schonherr2018adversarial,
  title={Adversarial attacks against automatic speech recognition systems via psychoacoustic hiding},
  author={Sch{\"o}nherr, Lea and Kohls, Katharina and Zeiler, Steffen and Holz, Thorsten and Kolossa, Dorothea},
  journal={arXiv preprint arXiv:1808.05665},
  year={2018}
}

@inproceedings{hou2025evaluating,
  title={Evaluating robustness of large audio language models to audio injection: An empirical study},
  author={Hou, Guanyu and He, Jiaming and Zhou, Yinhang and Guo, Ji and Qiao, Yitong and Zhang, Rui and Jiang, Wenbo},
  booktitle={Proceedings of the 2025 Conference on Empirical Methods in Natural Language Processing},
  pages={25671--25687},
  year={2025}
}

@article{cheng2025jailbreak,
  title={Jailbreak-audiobench: In-depth evaluation and analysis of jailbreak threats for large audio language models},
  author={Cheng, Hao and Xiao, Erjia and Shao, Jing and Wang, Yichi and Yang, Le and Shen, Chao and Torr, Philip and Gu, Jindong and Xu, Renjing},
  journal={arXiv preprint arXiv:2501.13772},
  year={2025}
}

@article{bai2022constitutional,
  title={Constitutional ai: Harmlessness from ai feedback, 2022},
  author={Bai, Yuntao and Kadavath, Saurav and Kundu, Sandipan and Askell, Amanda and Kernion, Jackson and Jones, Andy and Chen, Anna and Goldie, Anna and Mirhoseini, Azalia and McKinnon, Cameron and others},
  journal={URL https://arxiv. org/abs/2212.08073},
  volume={2212},
  year={2022}
}

@article{hu2024large,
  title={Large language models are efficient learners of noise-robust speech recognition},
  author={Hu, Yuchen and Chen, Chen and Yang, Chao-Han Huck and Li, Ruizhe and Zhang, Chao and Chen, Pin-Yu and Chng, EnSiong},
  journal={arXiv preprint arXiv:2401.10446},
  year={2024}
}

@article{rubenstein2023audiopalm,
  title={Audiopalm: A large language model that can speak and listen},
  author={Rubenstein, Paul K and Asawaroengchai, Chulayuth and Nguyen, Duc Dung and Bapna, Ankur and Borsos, Zal{\'a}n and Quitry, F{\'e}lix de Chaumont and Chen, Peter and Badawy, Dalia El and Han, Wei and Kharitonov, Eugene and others},
  journal={arXiv preprint arXiv:2306.12925},
  year={2023}
}

@inproceedings{wu2023decoder,
  title={On decoder-only architecture for speech-to-text and large language model integration},
  author={Wu, Jian and Gaur, Yashesh and Chen, Zhuo and Zhou, Long and Zhu, Yimeng and Wang, Tianrui and Li, Jinyu and Liu, Shujie and Ren, Bo and Liu, Linquan and others},
  booktitle={2023 IEEE automatic speech recognition and understanding workshop (ASRU)},
  pages={1--8},
  year={2023},
  organization={IEEE}
}

@inproceedings{yang2025audio,
  title={Audio is the achilles’ heel: Red teaming audio large multimodal models},
  author={Yang, Hao and Qu, Lizhen and Shareghi, Ehsan and Haffari, Gholamreza},
  booktitle={Proceedings of the 2025 Conference of the Nations of the Americas Chapter of the Association for Computational Linguistics: Human Language Technologies (Volume 1: Long Papers)},
  pages={9292--9306},
  year={2025}
}

@article{cesarini2024reverb,
  title={Reverb and noise as real-world effects in speech recognition models: a study and a proposal of a feature set},
  author={Cesarini, Valerio and Costantini, Giovanni},
  journal={Applied Sciences},
  volume={14},
  number={23},
  pages={11446},
  year={2024},
  publisher={MDPI}
}

@article{chowdhury2024breaking,
  title={Breaking down the defenses: A comparative survey of attacks on large language models},
  author={Chowdhury, Arijit Ghosh and Islam, Md Mofijul and Kumar, Vaibhav and Shezan, Faysal Hossain and Jain, Vinija and Chadha, Aman},
  journal={arXiv preprint arXiv:2403.04786},
  year={2024}
}

@article{jabbar2025red,
  title={Red teaming large language models: A comprehensive review and critical analysis},
  author={Jabbar, Muhammad Shahid and Al-Azani, Sadam and Alotaibi, Abrar and Ahmed, Moataz},
  journal={Information Processing \& Management},
  volume={62},
  number={6},
  pages={104239},
  year={2025},
  publisher={Elsevier}
}

@inproceedings{djanibekov2025spirit,
  title={Spirit: Patching speech language models against jailbreak attacks},
  author={Djanibekov, Amirbek and Mukhituly, Nurdaulet and Inui, Kentaro and Aldarmaki, Hanan and Lukas, Nils},
  booktitle={Proceedings of the 2025 Conference on Empirical Methods in Natural Language Processing},
  pages={14514--14531},
  year={2025}
}

@article{alzantot2018did,
  title={Did you hear that? adversarial examples against automatic speech recognition},
  author={Alzantot, Moustafa and Balaji, Bharathan and Srivastava, Mani},
  journal={arXiv preprint arXiv:1801.00554},
  year={2018}
}

@article{deng2016deep,
  title={Deep learning: from speech recognition to language and multimodal processing},
  author={Deng, Li},
  journal={APSIPA Transactions on Signal and Information Processing},
  volume={5},
  pages={e1},
  year={2016},
  publisher={Cambridge University Press}
}

@inproceedings{maas2012recurrent,
  title={Recurrent neural networks for noise reduction in robust asr.},
  author={Maas, Andrew L and Le, Quoc V and O'neil, Tyler M and Vinyals, Oriol and Nguyen, Patrick and Ng, Andrew Y},
  booktitle={Interspeech},
  volume={2012},
  pages={22--25},
  year={2012}
}

@inproceedings{o2021conformer,
  title={A conformer-based asr frontend for joint acoustic echo cancellation, speech enhancement and speech separation},
  author={O'Malley, Tom and Narayanan, Arun and Wang, Quan and Park, Alex and Walker, James and Howard, Nathan},
  booktitle={2021 IEEE Automatic Speech Recognition and Understanding Workshop (ASRU)},
  pages={304--311},
  year={2021},
  organization={IEEE}
}

@article{yao2024minicpm,
  title={Minicpm-v: A gpt-4v level mllm on your phone},
  author={Yao, Yuan and Yu, Tianyu and Zhang, Ao and Wang, Chongyi and Cui, Junbo and Zhu, Hongji and Cai, Tianchi and Li, Haoyu and Zhao, Weilin and He, Zhihui and others},
  journal={arXiv preprint arXiv:2408.01800},
  year={2024}
}

@article{chu2023qwen,
  title={Qwen-audio: Advancing universal audio understanding via unified large-scale audio-language models},
  author={Chu, Yunfei and Xu, Jin and Zhou, Xiaohuan and Yang, Qian and Zhang, Shiliang and Yan, Zhijie and Zhou, Chang and Zhou, Jingren},
  journal={arXiv preprint arXiv:2311.07919},
  year={2023}
}

@inproceedings{held2025distilling,
  title={Distilling an end-to-end voice assistant without instruction training data},
  author={Held, William and Zhang, Yanzhe and Li, Minzhi and Shi, Weiyan and Ryan, Michael J and Yang, Diyi},
  booktitle={Proceedings of the 63rd Annual Meeting of the Association for Computational Linguistics (Volume 1: Long Papers)},
  pages={7876--7891},
  year={2025}
}

@article{hughes2024best,
  title={Best-of-n jailbreaking},
  author={Hughes, John and Price, Sara and Lynch, Aengus and Schaeffer, Rylan and Barez, Fazl and Koyejo, Sanmi and Sleight, Henry and Jones, Erik and Perez, Ethan and Sharma, Mrinank},
  journal={arXiv preprint arXiv:2412.03556},
  year={2024}
}

@inproceedings{lv2025adappa,
  title={Adappa: Adaptive position pre-fill jailbreak attack approach targeting llms},
  author={Lv, Lijia and Zhang, Weigang and Tang, Xuehai and Wen, Jie and Liu, Feng and Han, Jizhong and Hu, Songlin},
  booktitle={ICASSP 2025-2025 IEEE International Conference on Acoustics, Speech and Signal Processing (ICASSP)},
  pages={1--5},
  year={2025},
  organization={IEEE}
}

@inproceedings{ren2024codeattack,
  title={Codeattack: Revealing safety generalization challenges of large language models via code completion},
  author={Ren, Qibing and Gao, Chang and Shao, Jing and Yan, Junchi and Tan, Xin and Lam, Wai and Ma, Lizhuang},
  booktitle={Findings of the Association for Computational Linguistics: ACL 2024},
  pages={11437--11452},
  year={2024}
}

@inproceedings{ao2022speecht5,
  title={Speecht5: Unified-modal encoder-decoder pre-training for spoken language processing},
  author={Ao, Junyi and Wang, Rui and Zhou, Long and Wang, Chengyi and Ren, Shuo and Wu, Yu and Liu, Shujie and Ko, Tom and Li, Qing and Zhang, Yu and others},
  booktitle={Proceedings of the 60th Annual Meeting of the Association for Computational Linguistics (Volume 1: Long Papers)},
  pages={5723--5738},
  year={2022}
}

@article{zou2023universal,
  title={Universal and transferable adversarial attacks on aligned language models},
  author={Zou, Andy and Wang, Zifan and Carlini, Nicholas and Nasr, Milad and Kolter, J Zico and Fredrikson, Matt},
  journal={arXiv preprint arXiv:2307.15043},
  year={2023}
}

\end{document}